\documentclass[manuscript]{aastex}
\usepackage{lscape}
\usepackage{graphicx}
\shortauthors{Kim et al.}
\received{19 July 2015}
\accepted{15 November 2015}

\begin{document}
\title{Near-Infrared Polarization Source Catalog of the Northeastern Regions of the Large Magellanic Cloud}

\author{Jaeyeong Kim\altaffilmark{1},
        Woong-Seob Jeong\altaffilmark{2}$^{,}$\altaffilmark{3}$^{,}$\altaffilmark{5}\footnotetext[5]{Corresponding author},
        Soojong Pak\altaffilmark{1}, Won-Kee Park\altaffilmark{2}, and Motohide Tamura\altaffilmark{4}}
\affil{$^1$ School of Space Research, Kyung Hee University,
            1 Seocheon-dong, Giheung-gu, Yongin, Gyeonggi-do 446-701,
            Republic of Korea; jaeyeong@khu.ac.kr}
\affil{$^2$ Korea Astronomy and Space Science Institute,
            776 Daedeok-daero, Yuseong-gu, Daejeon 305-348,
            Republic of Korea; jeongws@kasi.re.kr}
\affil{$^3$ Korea University of Science and Technology,
            217 Gajeong-ro, Yuseong-gu, Daejeon 305-350,
            Republic of Korea}
\affil{$^4$ The University of Tokyo / National Astronomical Observatory of Japan / Astrobiology Center,
            2-21-1 Osawa, Mitaka, Tokyo 181-8588,
            Japan}

\begin{abstract}
We present a near-infrared band-merged photometric and polarimetric catalog for the 39$\arcmin$ $\times$ 69$\arcmin$ fields on the northeastern part of the Large Magellanic Cloud (LMC), which were observed using SIRPOL, an imaging polarimeter of the InfraRed Survey Facility (IRSF). This catalog lists 1,858 sources brighter than 14 mag at $H$ band with polarization signal-to-noise ratio greater than three in the $J$, $H$, or $K_s$ bands. Based on the relationship between the extinction and the polarization degree, we argue that the polarization mostly arises from dichroic extinctions caused by local interstellar dust in the LMC. This catalog allows us to map polarization structures to examine the global geometry of the local magnetic field, and to show a statistical analysis of polarization of each field to understand its polarization properties. At the selected fields with coherent polarization position angles, we estimate magnetic field strengths in the range of 3$-$25 $\mu$G using the Chandrasekhar-Fermi method. This implies the presence of large-scale magnetic fields on a scale of around one hundred parsecs. When comparing mid and far-infrared dust emission maps, we confirmed that the polarization patterns are well aligned with molecular clouds around the star-forming regions.
\end{abstract}


\section{Introduction}
Magnetic fields are important for understanding galactic dynamics and star formation processes. To date, various methods for measuring interstellar magnetic fields have been devised to probe the star-forming regions in our Galaxy and in external galaxies \citep{vrb76,dav51,laz07}. When background starlight passes through dust grains aligned with the local magnetic field, the light is linearly polarized by dichroic extinction parallel to the grain's long axis.

The Large Magellanic Cloud (LMC) is a unique target for studying magnetic fields and star-forming processes. Due to its proximity and face-on orientation, large-scale interactions with the halo of the Milky Way and the Small Magellanic Cloud (SMC), and small-scale structures of the local star-forming regions, can be traced \citep{kim98,pak98,kim11}. Previous studies of magnetic fields in the LMC were carried out using multi-frequency surveys of diffuse synchrotron emission in radio bands \citep{way90,hay91,gae05,mao12}. Optical polarization \citep{vis66,sch70,sch76,mat70} have shown the existence of large-scale magnetic fields in the LMC. However, these results were restricted to only bright stars. \citet{wis07} (hereafter W07) also showed the detailed polarization map of NGC 2100 on the eastern side of 30 Doradus and its surrounding regions, in optical bands. The near-infrared band has an advantage over the optical band. This is because the background starlight in star-forming regions suffers heavy extinction \citep{tam87,tam88,sat88}. \citet{nak07} (hereafter N07) studied magnetic field structures 7.7$\arcmin$ $\times$ 7.7$\arcmin$ around 30 Doradus using near-IR polarimetry. They revealed more detailed magnetic field structures that are aligned roughly in the east-west direction compared with the previous results in the optical band \citep{vis66,sch70,mat70}. These field structures are associated with expanding shells around 30 Doradus.

We carried out a near-IR photometric and polarimetric study for the 39$\arcmin$ $\times$ 69$\arcmin$ fields in the northeastern part of the LMC. The target fields contain four important star-forming regions, 30 Doradus, N158, N159 and N160. The initial investigation of polarimetric data around 30 Doradus was presented in \citet{kim11}. However, the survey was limited to a 20$\arcmin$ $\times$ 20$\arcmin$ region and the data were hampered by unstable weather conditions. In this paper, we analyzed additional data sets for more regions. The procedures to obtain the photometric results are described in Section 2. We compiled photometric and polarimetric data for each observation field in a catalog. In Section 3, we examined the origin of the polarization, based on the wavelength dependence. In Section 4, we showed polarization vector maps for each observation field and discussed the related magnetic field structures. We also calculated magnetic field strengths in five sample fields. Finally, we discussed the correlation of our polarization results with dust emission and CO distribution.

\section{Observations and Data Reduction}

\subsection{Observational information}
\begin{figure*}[p]
\centering
\includegraphics[scale=1]{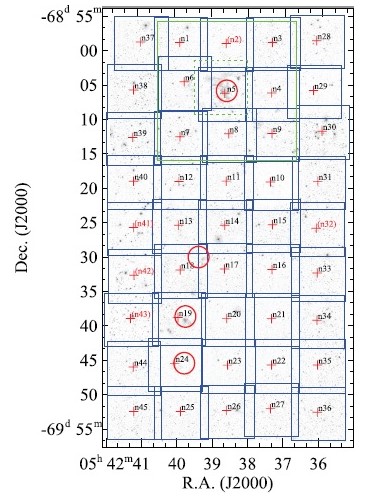}
\caption{The observed fields in the northeastern regions of the LMC. The figure shows total 5$\times$9 fields observed with SIRPOL on 2008 December 25, 26, 27, and 30, and 2011 December 10 and 11. Red open circles represent star-forming regions: 30 Doradus, N158, N160, and N159, from north to south direction. The field center and area (7.7$\arcmin$ $\times$ 7.7$\arcmin$) are displayed as red cross and blue box, respectively. The fields of view covered by N07 and \citet{kim11} are also highlighted as green dotted and solid boxes, respectively. Field numbers in parentheses denote the discarded fields due to unstable weather conditions. }\label{fig1}
\end{figure*}

We observed the northeastern regions in the LMC with the infrared camera SIRIUS \citep{nag03} and the polarimeter SIRPOL \citep{kan06} at the Infrared Survey Facility (IRSF) 1.4 m telescope at the South African Astronomical Observatory (SAAO) on 2008 December 25$-$30 and 2011 December 2$-$11. This system has a field of view (7.7$\arcmin$ $\times$ 7.7$\arcmin$) and a pixel scale of 0.45$\arcsec$ pixel$^{-1}$. One data set for a target field consists of 20-second exposures at 10 dithered positions for four wave-plate angles (0$\arcdeg$, 45$\arcdeg$, 22.5$\arcdeg$, and 67.5$\arcdeg$) in the \textit{J} (1.25 $\mu$m), \textit{H} (1.63 $\mu$m), and \textit{$K_s$} (2.14 $\mu$m) bands. The target regions of 39$\arcmin$ $\times$ 69$\arcmin$ are tiled with 45 fields with grid spacing of 6.5$\arcmin$ $\times$ 6.5$\arcmin$ (Figure~\ref{fig1}). We list the observation log in Table~\ref{tbl-1}.

\subsection{Data reduction procedure}
We used the SIRPOL data reduction pipeline \citep{kan06} at the IRAF\footnote{IRAF is distributed by the US National Optical Astronomy Observatories, which are operated by the Association of Universities for Research in Astronomy, Inc., under cooperative agreement with the National Science Foundation.}(Image Reduction \& Analysis Facility). The pipeline incorporates the following procedures: flat field correction, sky subtraction, and combining of dithered frames. We checked the output magnitude variations in 40 sequential frames to filter bad data affected by non-photometric weathers. The fields with magnitude variation larger than 0.03 mag were rejected (see Figure~\ref{fig1}).

We used Source Extractor (SExtractor) for source detection and aperture photometry \citep{ber96}. Various SExtractor parameters were optimized (i.e., detection threshold of 5 $\sigma$, background box size of 16 pixels, background filter of 3 $\times$ 3 meshes, and aperture diameter of 8 pixels). We measured the instrumental magnitudes of stars for four wave-plate angles in the $J$, $H$, and $K_s$ bands. The pixel coordinates of the sources were converted to celestial coordinates using the 2MASS All Sky Point Source Catalogue \citep{skr06}. The instrumental magnitude and color were calibrated using the equations below:\\
\begin{equation}
 M_{2MASS} = M_{IRSF} + \alpha_{1} \times C_{IRSF} + \beta_{1},\\\end{equation}
\begin{equation}
 C_{2MASS} = \alpha_{2} \times C_{IRSF} + \beta_{2},\\\end{equation}
where $M_{IRSF}$ and $C_{IRSF}$ are the instrumental magnitudes and the colors from the Stokes $I$ images, and $M_{2MASS}$ and $C_{2MASS}$ are the magnitude and the color in the 2MASS system. The transformation coefficients, $\alpha$ and $\beta$, for each field were estimated using a robust least absolute deviation method.

We calculated polarization of point sources using Stokes parameters $I$, $Q$, and $U$ as described by \citet{kim11}. Flux errors of sources at each wave-plate angle (0$\arcdeg$, 45$\arcdeg$, 22.5$\arcdeg$, and 67.5$\arcdeg$) are described as $\sigma_{I_{1}}$, $\sigma_{I_{2}}$, $\sigma_{I_{3}}$, and $\sigma_{I_{4}}$, respectively. The uncertainty for $P$ (\%) was derived as follows:\\
\begin{equation}
\overline{\sigma}_{Q} = \sqrt{\left ( \frac{\sigma_{Q}}{I}\right )^{2} +\left ( \frac{Q}{I^{2}}\times \sigma_{I} \right )^{2}},\\\end{equation}
\begin{equation}
\overline{\sigma}_{U} = \sqrt{\left ( \frac{\sigma_{U}}{I}\right )^{2} +\left ( \frac{U}{I^{2}}\times \sigma_{I} \right )^{2}},\\\end{equation}
\begin{equation}
\sigma_{P} = \frac{\sqrt{\left ( \frac{Q}{I} \right )^{2}\times \overline{\sigma}_{Q}^{2}+\left ( \frac{U}{I} \right )^{2}\times \overline{\sigma}_{ U}^{2}}}{P}, \\\end{equation}
where $\overline{\sigma}_{Q}$ and $\overline{\sigma}_{U}$ are normalized errors with $I$. The errors of Stokes parameters $\sigma_{I}$, $\sigma_{Q}$, and $\sigma_{U}$ can be written as $\sqrt{\sigma_{Q}^{2}+\sigma_{U}^{2}}/{2}$, $\sqrt{\sigma_{I_{1}}^{2}+\sigma_{I_{2}}^{2}}$, and $\sqrt{\sigma_{I_{3}}^{2}+\sigma_{I_{4}}^{2}}$, respectively.

\section{Results}

\subsection{Estimation of the photometric and the polarimetric accuracy}
\begin{figure*}[p]
\centering
\includegraphics[scale=1]{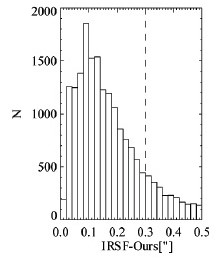}
\caption{Histogram of position offsets between the IRSF catalog and our sources. Distribution peak is around 0.1$\arcsec$. The vertical dashed line indicates our matching criterion for the catalog.}\label{fig2}
\end{figure*}

\begin{figure*}[p]
\centering
\includegraphics[scale=1]{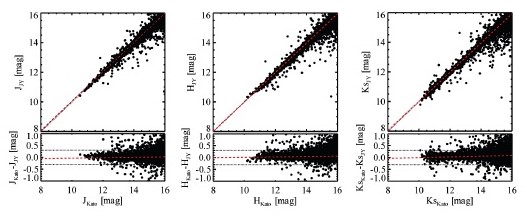}
\caption{Magnitude differences of sources between IRSF catalog and ours as a function of magnitude at the $J$ (left), $H$ (middle), and $K_S$ (right) bands. Black dotted lines in lower panels indicate the magnitude difference of 0.3 mag. All red dashed lines indicate results of the least square fitting for the sources.}\label{fig3}
\end{figure*}

A total of 25,488 sources were detected with the SExtractor. To verify the photometric accuracy, we compared our data set with the IRSF Magellanic Cloud point source catalog (hereafter the IRSF catalog) in \citet{kat07}. The IRSF catalog provides photometry results in the $J$, $H$, and $K_s$ bands. Figure~\ref{fig2} shows a histogram of position offsets of our sources that matched theirs. Most sources have the offset around 0.1$\arcsec$ and we retain only the sources with offset less than $0.3 \arcsec$. The proper motion effect between two data sets is negligible: $\mu_{\alpha}\cos\delta$ = 1.89 $\pm$ 0.27 mas yr$^{-1}$, and $\mu_{\delta}$ = 0.39 $\pm$ 0.27 mas yr$^{-1}$ \citep{vie10}. The magnitude differences between the IRSF catalog and ours are shown in Figure~\ref{fig3}. Most sources brighter than 14 mag have differences in magnitude from the IRSF catalog less than 0.3 mag in the $J$, $H$, and $K_s$ bands.

\begin{figure*}[p]
\centering
\includegraphics[scale=1]{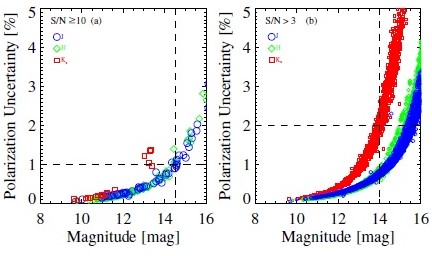}
\caption{Polarization uncertainty as a function of magnitude at the $J$ ($circle$), $H$ ($diamond$), and $K_S$ ($box$) bands. (a) The sources with polarization signal-to-noise ratio $P/\sigma_P$ $\geq$ 10 are included in the figure. Under 14.5 magnitude (vertical dashed line), most sources have the uncertainty of polarization degree smaller than 1\% (horizontal dashed line). (b) The sources with polarization signal-to-noise ratio $P/\sigma_P$ $>$ 3 are included in the figure. Under 14 magnitude (vertical dashed line), most sources have the uncertainty of polarization degree smaller than 2\% (horizontal dashed line).}\label{fig4}
\end{figure*}

\begin{figure*}[p]
\centering
\includegraphics[scale=1]{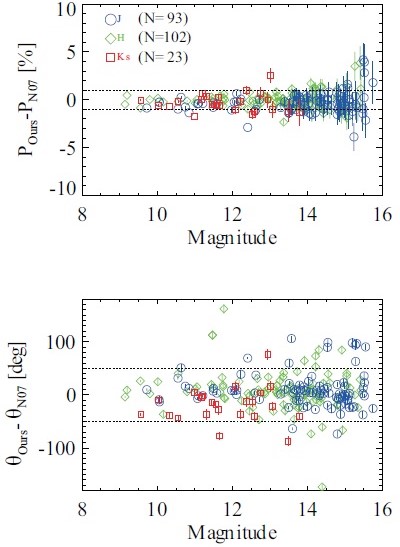}
\caption{Polarization degree (top) and position angle (bottom) differences of sources matched between N07 and ours in 30 Doradus as a function of magnitudes at the $J$ ($circle$), $H$ ($diamond$), and $K_S$ ($box$) bands. Scattered ranges of the sources brighter than 14 mag at polarization degree and position angle are denoted by dotted lines.}\label{fig5}
\end{figure*}

The polarization uncertainty with magnitude is shown in Figure~\ref{fig4}. Among the sources with $P/\sigma_P$ (polarization signal-to-noise ratio) greater than 10, the polarization uncertainty is lower than 1\% for sources brighter than 14.5 mag (Figure~\ref{fig4} a). We also compared polarimetric results with previous studies. N07 had observed the central 7.7$\arcmin$ $\times$ 7.7$\arcmin$ (one field) region of 30 Doradus using the same SIRIUS/SIRPOL system. The integration time of 1,480 seconds per wave-plate angle makes N07 a very good reference for verifying the polarimetric accuracy of our data. We examined the polarimetric results for $P/\sigma_P$ $>$ 3 between N07 and our 30 Doradus field, and Figure~\ref{fig5} shows the comparison results. The difference of polarization degree increases at around 14 mag in the $J$ and $H$ bands. Though the difference in the polarization angle shows some notable scatter over the entire range of magnitude, the scatter is quite constant down to 14 mag in the $J$ and $H$ bands.

\begin{figure*}[p]
\centering
\includegraphics[scale=1]{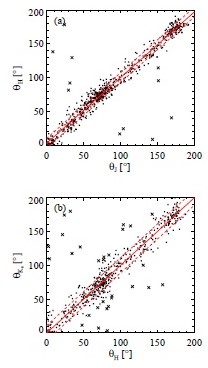}
\caption{Polarization position angle difference with (a) $J$ versus $H$ band and (b) $H$ versus $K_s$ band for 697 and 508 sources, respectively. These sources are selected from the following criteria: m$_{H}$ $<$ 14 mag and $P/\sigma_P$ $>$ 3. Both figures show similar distribution for polarization position angles within one standard deviation (red dashed lines) from the correlation slope (red solid line) of (a) and (b), respectively. The cross symbol shows a rejected source that a deviation of polarization position angle from the fitted line is greater than 30$^\circ$. Most of sources in our catalog have the uncertainty of polarization position angles smaller than 10$^\circ$.}\label{fig6}
\end{figure*}

As a result, we adopted all sources brighter than 14 mag in the $H$ band. In order to increase the number of meaningful sources, we extended the criteria of $P/\sigma_P$ to include $'$3$'$. Most sources in Figure~\ref{fig4} (b) have their polarization uncertainties $\leq$ 2\% for sources brighter than 14 mag. Typical polarization uncertainties between 13.9 and 14.1 mag in $H$ band are 1.12 $\pm$ 0.25, 0.74 $\pm$ 0.06, and 1.68 $\pm$ 0.27 $\%$ in the $J$, $H$, and $K_s$ bands, respectively. We checked the accuracy of polarization position angles for the selected data by comparing differences in the polarization position angles among the three bands. The upper and lower panels of Figure~\ref{fig6} show the distribution of polarization position angles for $J$ versus $H$ band, and $H$ versus $K_s$ band, respectively. The estimated uncertainty of the polarization position angles is smaller than 10$^\circ$. As seen in Figure~\ref{fig6}, we confirmed that most of the sources in our catalog show consistent polarization position angles (within one standard deviation) from the correlation slope.

\subsection{Catalog}
The final catalog includes all reliable sources. Among 25,488 detected sources, 1,858 sources were selected based on the following criteria: m$_{H}$ $<$ 14 mag and $P/\sigma_P$ $>$ 3 for at least one band, where $\sigma_P$ indicates the polarization uncertainty. Table~\ref{tbl-2} lists photometric and polarimetric results for the compiled catalog sources. In addition, we compiled proper motion data from Southern Proper Motion (SPM) material \citep{vie10}. Information for each column of the table is given below:\\
Column (1) Observation field names that include observation date;\\
Column (2) Source ID; the formats referred to the equatorial coordinates;\\
Column (3)$-$(4) Equatorial coordinates (J2000.0) in decimal degrees;\\
Column (5)$-$(10) $J$, $H$, and $K_{s}$ magnitude and error;\\
Column (11)$-$(16) $J$, $H$, and $K_{s}$ polarization degree and error in units of percentage;\\
Column (17)$-$(22) $J$, $H$, and $K_{s}$ polarization position angle and error in units of degrees;\\
Column (23)$-$(24) Absolute proper motion in right ascension and error in units of mas yr$^{-1}$ from SPM catalog;\\
Column (25)$-$(26) Absolute proper motion in declination and error in units of mas yr$^{-1}$ from SPM catalog;\\
Column (27)$-$(28) Johnson B and V magnitude from SPM catalog.

\subsection{Polarization relationship between $J$, $H$, and $K_s$ bands}
\begin{figure*}[p]
\centering
\includegraphics[scale=1]{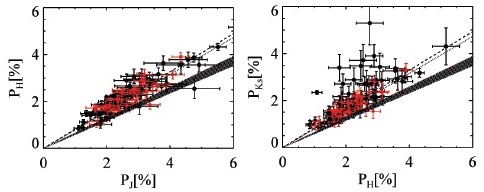}
\caption{$P_{J}$ vs. $P_{H}$ (left) and $P_{H}$ vs. $P_{K_{s}}$ (right) for the selected 106 sources with $P/\sigma_P$ $>$ 5 and $\sigma_P$ $\leq$ 1 in all bands. The best-fit values for the slopes are $P_{H}/P_{J}$ = 0.81 and $P_{K_{s}}/P_{H}$ = 0.82 (dashed lines). The hatched area shows previous studies \citep{nagt90,mar90,mar92}, and black dotted line is the best-fit value from \citet{nak07}. The red symbol shows the sources matched with AKARI LMC catalog.}\label{fig7}
\end{figure*}

\citet{ser75} first formulated the wavelength dependence of interstellar polarization in the Galaxy, and \citet{whi78} modified the empirical relationship between the wavelength and the size of the dust grains. The modified "Serkowski law" was represented by a power-law dependence of polarization as below:\\
\begin{equation}
P_{\lambda} \propto \lambda^{-\beta}\\\end{equation}
with $\beta$ of 1.6$-$2.0 in the near-IR (1.25$\leq\lambda\leq$2.2) in previous studies \citep{nagt90,mar90,mar92}. For the 30 Doradus region, N07 presented a slightly different power-law index of $\beta$ $=$ 0.9. As seen in Figure~\ref{fig7}, our result is consistent with that of N07. For your reference, we also plotted the sources matched with the AKARI LMC point-source catalog to check the possibility of intrinsic polarization (see Section \ref{sec:AKARI_MIR} for details).

\begin{figure*}[p]
\centering
\includegraphics[scale=1]{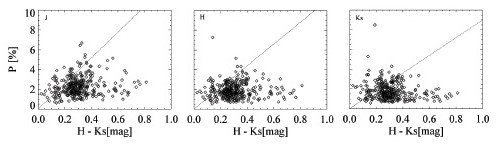}
\caption{Degree of polarization $P$ for the $J$(left), $H$(middle), and $K_{s}$(right) bands vs. $H-K_{s}$ color. Dotted lines are empirical upper limits, $P_{max}$ \citep{jon89}. Open diamonds are sources with $P/\sigma_P$ $>$ 3 and $\sigma_P$ $\leq$ 1 in all bands.}\label{fig8}
\end{figure*}

We examined the correlation between the observed polarization efficiency and the upper limit of polarization degree. Figure~\ref{fig8} shows the result of polarization efficiency for the sources with $P/\sigma_P$ $>$ 3 and $\sigma_P$ $\leq$ 1 in all bands. Most of the sources are located below the upper limit of interstellar polarization (dashed line). This means that the observed polarization in the LMC is of interstellar origin.

\subsection{Possibility of intrinsic polarization component}\label{sec:AKARI_MIR}
$AGB$ stars are often associated with nebulosity and therefore the polarization might be intrinsic rather than interstellar. We examine this possibility from mid-infrared observation data of the AKARI LMC point-source catalog. Using a color-color diagram and a color-magnitude diagram of the five photometric bands at 3.2, 7, 11, 15, and 24 $\mu$m, \citet{kat12} classified the sources with dusty C-rich and O-rich $AGB$ stars.

In our catalog, 208 sources have been matched to the AKARI LMC point sources in the 3.2, 7, and 11 $\mu$m bands. The matched AKARI sources were divided into $AGB$ stars without dusty envelope and dusty C-rich/O-rich $AGB$ stars. However, distributions of the polarization degrees and position angles do not show any significant differences between two groups. In addition, the matched AKARI sources do not show any different power-law relationship (Figure~\ref{fig7}). This implies that the detected polarization is of interstellar origin rather than being intrinsic to $AGB$ stars.

\section{Discussion}

\subsection{Polarization results}
\begin{figure*}[p]
\centering
\includegraphics[scale=1]{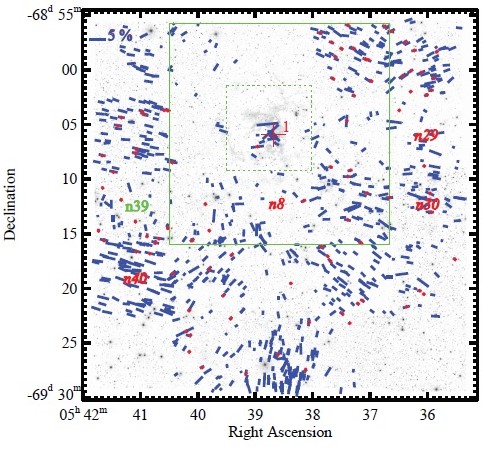}
\caption{Polarization vector map of the $H$ band for the region around 30 Doradus. Stars which have $P/\sigma_P$ $>$ 3 are displayed with blue lines. The length of the vectors denotes the degree of polarization, and its scale is shown in upper left corner. One field overlapped with W07 and other three sample fields are denoted by the field names in red color and bold. Those fields were used for the calculation of the magnetic field strength. The fields of view covered by N07 and \citet{kim11} are also highlighted as green dotted and solid boxes, respectively. The location of the optical polarimetry by W07 is eastern side of $n39$ field (green-colored field name). Red diamonds are the AKARI sources which we matched in the section \ref{sec:AKARI_MIR}. The center position of the 30 Doradus are denoted by the red-colored cross with a number of 1.}\label{fig9}
\end{figure*}

\begin{figure*}[p]
\centering
\includegraphics[scale=1]{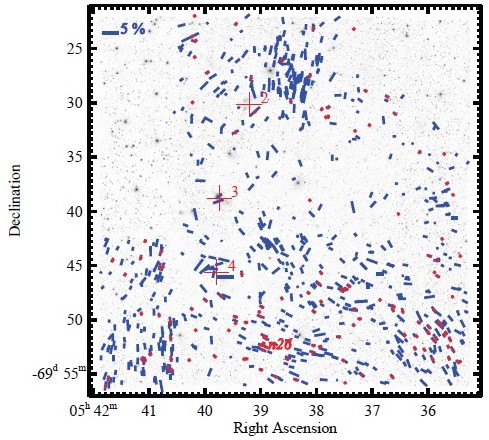}
\caption{Same with Figure~\ref{fig9} but for regions around N158, N160, N159, and southern field. The center positions of N158, N160, and N159 are denoted by the red-colored crosses with numbers of 2 to 4, respectively. One sample field to calculate the magnetic field strength is denoted by the field name in red color and bold.}\label{fig10}
\end{figure*}

Figures~\ref{fig9} and ~\ref{fig10} show the $H$-band polarization maps in the observed LMC fields. Each vector represents a point source with $P/\sigma_P$ $>$ 3. The number of selected stars was 875, 1,468, and 732, in the $J$, $H$, and $K_s$ bands, respectively. Most of these vectors are directed either from north to south or from northeast to southwest.

In the region around 30 Doradus (Figure~\ref{fig9}), most of the polarization vectors have directions from northeast to southwest. In the southeastern region from 30 Doradus, the distribution of polarization vectors shows a shell-like structure encircling $n39$ field. W07 presented the magnetic field structure of NGC 2100, of which the western part overlapped our fields. The polarization feature in the region of overlap shows similar polarization vector patterns. W07 reported that the origin of the magnetic field geometry around NGC 2100 includes massive outflows moving eastward from the 30 Doradus region. The coherent pattern of the polarization vectors in the north and south of $n39$ field is consistent with that of W07. They reported that the pattern is influenced by magnetic field lines along the eastward outflows. In addition, other patterns in $n39$ field can be explained by interaction between the environment of NGC 2100 and the eastward outflows. The large-scale outflow from 30 Doradus is one plausible source of the influence on the structural formation of the supergiant shell on the western side of 30 Doradus (LMC 3 region). Overall vectors in the western regions of 30 Doradus show coherent patterns toward the direction of the LMC 3 region identified in previous studies \citep{mea80,poi01,bcg08,daw13}. As suggested by W07, the extended uniform patterns of a magnetic field can be a clue to large-scale outflows toward the boundaries of a supergiant shell.

The polarization vectors in Figure~\ref{fig10} show complex distribution. At the eastern part, most of the polarization vectors are directed north-south. However, vectors at the other side show a different distribution structure, more like a tilted S-shape. We suspect that these patterns of polarization vectors are related to the magnetic field lines south of 30 Doradus. Previous studies \citep{way90,hay91,sk92} of the geometry of the magnetic field south of 30 Doradus suggested that it is aligned with the filamentary features B and C of Figure 2 in \citet{fei87}. Two patterns show direction similar to those of these filamentary features. \citet{way90} and \citet{mao12} also proposed that the gaseous H\textsc{i} spiral features are likely associated with the large-scale magnetic fields toward the south direction.

A statistical analysis was conducted field-by-field to understand the interstellar polarization properties. In Table~\ref{tbl-3}, the number of stars with $P/\sigma_P$ $>$ 3 for each band is given in columns (2$-$4). The average polarization degrees for each band is given in columns (5), (7), and (9), and their mean uncertainties in column (6), (8), and (10). The average polarization position angles at each band is given in columns (11), (13), and (15) with their standard deviation in columns (12), (14), and (16). We examined the magnetic field strength of the fields overlapped with those described by W07 and other sample fields using the analysis of \citet{chf53}, which is described by
\begin{equation}
B = Q\sqrt{4\pi\rho}\frac{\delta\textit{v}_{los}}{\delta\theta }
\end{equation}
where $\rho$ is the mean density, $\delta\textit{v}_{los}$ is the velocity dispersion in the line-of-sight, and $\delta\theta$ is the dispersion in the Gaussian fit of polarization position angles. \citet{kim07} presented a 21 cm neutral hydrogen interferometric survey of the LMC in a catalog of H\textsc{i} gas clumps or clouds with 16, 32, and 64 K brightness temperature thresholds. By matching their catalog locations with positions of our observed fields, we obtained $\delta\textit{v}_{los}$ from their survey results. The applied $\delta\textit{v}_{los}$ of the matched sample fields are tabulated in Table~\ref{tbl-4}. As \citet{cru04} reported, the value of a factor of order unity, Q, shows the best result when the value from \citet{ost01} is adopted. By simulation, \citet{ost01} determined that Q is approximately 0.46$-$0.51, when the observed dispersions in polarization position angles indicate relatively strong magnetic field strengths ($\delta\theta$ $<$ 25$\arcdeg$). We set the value of Q to be 0.5, as suggested by \citet{cru04}. We used column (14) in Table~\ref{tbl-3} to select sample fields showing coherent distribution of polarization position angles. For the coherent distribution, $\delta\theta$ was calculated by a Gaussian fit centered at average polarization position angles in Table~\ref{tbl-3}. We tested whether the distribution of polarization position angles is well fitted to a Gaussian distribution by applying the Martinez-Iglewicz normality test \citep{mar81},
\begin{equation}
I= \frac{\sum_{i=1}^{n}\left ( \theta_{i} -\langle\theta_{H}\rangle \right )^{2}}{\left (n-1  \right )\delta\theta ^{2}} \end{equation}
where $I$ is the Martinez-Iglewicz test statistic, $\langle\theta_{H}\rangle$ is average polarization position angles in the $H$ band. If the distribution possesses significant deviation from normality, it indicates the presence of possible sub-structure. To eliminate these cases, we selected fields showing test value $I$ close to $'$1$'$ (Table~\ref{tbl-4}). W07 used H\textsc{i} number density from \citet{poi99} to estimate mean density in their observed fields. We also assumed the number density of 4 cm$^{-3}$ as used in W07. As seen in Table~\ref{tbl-4}, we tabulated the magnetic field strengths for the five sample fields. One of them is the same region considered in W07, and it shows a similar magnetic field strength. The derived magnetic field strengths and sizes of the selected regions are similar to the properties of a typical cloud complex (i.e., scale 10$-$100 parsecs and magnetic field strength ranging from 10 to 30 $\mu$G) as reported by \citet{cha11}.

\subsection{Polarization structure with molecular cloud studies}\label{sec:Pol_structure}
In order to understand the pattern of polarization vectors, we examined our $H$-band polarization map by comparing the literature on mid to far-infrared dust emission maps and CO gas emission maps. We assume that the position angle of polarization indicates the direction of the magnetic field, because the observed polarization originated from interstellar dust grains with short axis aligned with the local magnetic field in the LMC.

\begin{figure*}[p]
\centering
\includegraphics[scale=1]{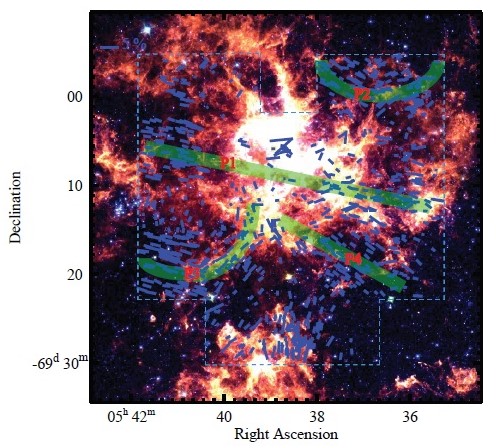}
\caption{Distribution of polarization vectors ($P$ $>$ 3$\sigma_P$) from $H$ band data overlaid on the color composite image of Spitzer IRAC 3.6 ($blue$), 5.8 ($green$), and 8 $\mu$m ($red$) bands in 30 Doradus region. Stars which $P/\sigma_P$ $>$ 3 are displayed in blue lines. Prominent patterns of polarization vectors are described as the green shaded curves (P1, P2, P3, and P4). The length of the lines denotes the degree of polarization which scale is shown in the upper left corner. The blue dashed box denotes the border of our data fields.}\label{fig11}
\end{figure*}

\begin{figure*}[p]
\centering
\includegraphics[scale=1]{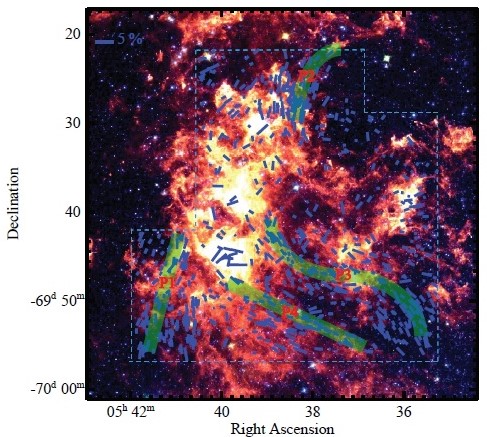}
\caption{Same with Figure~\ref{fig11} but for regions of N158, N160, N159, and southern field.}\label{fig12}
\end{figure*}

The IRAC data (3.6, 5.8, and 8 $\mu$m) from the Spitzer SAGE \citep{meix06} were compiled to trace dust emission features in the LMC fields more clearly. Although we have done a comparison with the Herschel data (100, 160, and 250 $\mu$m) from the HERITAGE project \citep{meix10}, we did not find any features different from those in the Spitzer data, due to relatively poor resolution. The IRAC data are displayed as a color composite image (shown in Figures~\ref{fig11} and~\ref{fig12}), and we overlaid our polarization vector map on these figures. The polarization vectors showing prominent patterns indicate the direction of the local magnetic field lines and we indicated these patterns with a green shaded curve in both Figure~\ref{fig11} and~\ref{fig12}.

The overall trend of the polarization vectors shows east-west direction across the 30 Doradus region (P1 in Figure~\ref{fig11}). Other distinctive patterns of polarization vectors are located at the outskirts field of 30 Doradus, associated with dust emission features (P2, P3, and P4 in Figure~\ref{fig11}). P2 is located at the northwestern boundary of the dust emission feature and it shows a U-shaped structure. N07 found a similar U-shaped structure of polarization vectors, and they proposed that expanding shells in 30 Doradus affected complex cloud structures and the associated magnetic field. P3 and P4 exhibit apparent patterns with similar direction to that of bright emission features.

Figure~\ref{fig12} shows the distribution of polarization vectors mainly following the molecular cloud ridge around star-forming regions of N158, N160, and N159. The vectors located at pattern P1 show distinct elongation extending southward, with direction similar to that of the dust emission structure. P2 also shows a uniform pattern directed north-south direction, but its shape is slightly curved along the dust emission feature. These patterns (P1 and P2) are regarded as magnetic field lines related to the extended feature of the molecular cloud ridge. However, polarization vectors located at the southwestern part of Figure~\ref{fig12} show patterns different from those of the eastern part. Polarization vectors in patterns P3 and P4 tend to be aligned in a tilted S-shape and east-west, respectively. We observed that the dust emission feature in the southwestern part of Figure~\ref{fig12} seems to be extended in this direction.

On the other hand, polarization vectors in the star-forming regions (central part of 30 Doradus in Figure~\ref{fig11} and bright emission features in Figure~\ref{fig12}) do not show any significant pattern. This is thought to be due to the turbulent and complex magnetic fields of the sub-structures in the star-forming region. To verify the structure of the expected turbulent magnetic fields, more detailed polarimetric data are required.

We also did a comparison with a velocity-integrated contour map of $^{12}$CO emission. NANTEN $^{12}$CO($J$=1-0) emissions have been detected in giant molecular cloud complexes in the LMC \citep{fuk08}. Although prominent $^{12}$CO emission features are located in the central 30 Doradus region, and in star-forming regions in the southern molecular cloud ridge, close correlation between magnetic fields and $^{12}$CO emission features was not found.

We concluded that the observed polarization vectors reflect a dust cloud structure resulting in polarization by dichroic extinction, and that they trace magnetic fields associated with dust clouds around star-forming region.

\section{Summary \& Conclusions}
We conducted near-IR imaging polarimetry for a large areal region covering ($\sim39\arcmin\times69\arcmin$) on the eastern side of the LMC. We made a band-merged catalog of photometric and polarimetric data in $J$, $H$, and $K_s$ bands. In this catalog, we compiled 1,858 stars in the region, brighter than 14 mag in $H$ band and $P/\sigma_P$ $>$ 3 for least one band. The magnitude, polarization degree and polarization position angle were listed in the catalog. In addition, we provided information on absolute proper motion and magnitudes of $B$ and $V$ matched from SPM data. Using this catalog, we obtained polarization vector maps and did a statistical analysis of polarization for the fields observed in the LMC. Previous similar catalogs did not cover such a large continuous area of the LMC (e.g., $\sim15\arcmin\times15\arcmin$ by W07), nor did they use the infrared range. Therefore, this catalog is a unique, up-to-date collection of polarization measurements for the LMC region.

The degree of polarization for the 106 sources in our catalog shows wavelength dependence similar to those reported by N07. We concluded that most of the sources in our catalog exhibit interstellar polarization, and that the dominant polarization mechanism is the result of dichroic extinction by dust grains aligned along magnetic fields, rather than due to intrinsic polarization of stars. Using the polarization vector maps, we traced the correlation of the polarization and the large-scale magnetic field structures in the observed regions. The geometry of the large-scale magnetic field structures around 30 Doradus, and of the southern star-forming regions, show relationships with the environment of nearby supergiant shells and the gaseous spiral features toward the south, respectively. The estimated magnetic field strengths for the selected fields (within one hundred parsecs) are in the range 3$-$25 $\mu$G, as determined by the Chandrasekhar-Fermi method. Judging from their size and magnetic field strength, those regions are regarded as a cloud complex associated with a nebula in the LMC. Prominent patterns of polarization vectors mainly follow dust emission features in mid to far-infrared bands, which implies that the large-scale magnetic fields are well involved in the structures of the dust cloud in the LMC, except for the dense star-forming regions. The cross matching of the data in this catalog, with that in other catalogs for in the LMC, will give useful information for probing the structure of magnetic fields and other astrophysical phenomena.

\acknowledgments
This work was supported by the National Research Foundation of Korea (NRF) grant, No. 2008-0060544, funded by the Korea government (MSIP).
We would like to thank Jungmi Kwon for giving a chance to study this work.
We would like to thank Prof. Shuji Sato for kindly providing comments to improve this paper.
This paper uses observations performed at the South African Astronomical Observatory.
This publication makes use of data products from the Two Micron All Sky Survey and observations with AKARI. The Two Micron All Sky Survey is a joint project of the University of Massachusetts and the Infrared Processing and Analysis Center/California Institute of Technology, funded by the National Aeronautics and Space Administration and the National Science Foundation. The AKARI is a JAXA project with the participation of ESA.
This research has made use of the NASA/ IPAC Infrared Science Archive, which is operated by the Jet Propulsion Laboratory, California Institute of Technology, under contract with the National Aeronautics and Space Administration.

{}

\clearpage

\begin{deluxetable}{cccccc}
\tabletypesize{\scriptsize}
\tablecaption{Observation log\label{tbl-1}}
\tablewidth{0pt}
\setlength{\tabcolsep}{0.02in}
\tablehead{\colhead{Field name} & \colhead{${{\alpha_{\circ\rm J2000.0}}}$} & \colhead{${\delta_{\circ\rm J2000.0}}$} & \colhead{Date (LT)}
& \colhead{Seeing in J}& \colhead{Magnitude variation} \\
\colhead{} & \colhead{} & \colhead{} & \colhead{} & \colhead{($\arcsec$)} & \colhead{(mag)}
}\startdata
n1  & 05 39 54.4 & -68 59 20.0 & 2008Dec30 & 1.70 & 0.022 \\
n3  & 05 37 25.6 & -68 59 19.9 & 2008Dec30 & 1.65 & 0.022 \\
n4  & 05 37 25.2 & -69 06 00.5 & 2008Dec25 & 1.58 & 0.023 \\
n5  & 05 38 40.1 & -69 06 00.0 & 2008Dec25 & 1.67 & 0.029 \\
n6  & 05 39 54.8 & -69 06 00.1 & 2008Dec25 & 1.68 & 0.024 \\
n7  & 05 39 55.1 & -69 12 40.2 & 2008Dec25 & 1.51 & 0.016 \\
n8  & 05 38 40.1 & -69 12 40.1 & 2008Dec25 & 1.54 & 0.022 \\
n9  & 05 37 25.0 & -69 12 39.9 & 2008Dec25 & 1.45 & 0.026 \\
n10 & 05 37 24.6 & -69 19 19.9 & 2008Dec25 & 1.91 & 0.023 \\
n11 & 05 38 40.0 & -69 19 20.2 & 2008Dec25 & 2.46 & 0.024 \\
n12 & 05 39 55.6 & -69 19 20.1 & 2008Dec25 & 1.62 & 0.023 \\
..  & .. .. .... & ... .. .... & 2011Dec11 & 1.48 & 0.024 \\
n13 & 05 39 56.0 & -69 26 00.4 & 2008Dec26 & 2.07 & 0.019 \\
..  & .. .. .... & ... .. .... & 2011Dec11 & 1.24 & 0.024 \\
n14 & 05 38 40.1 & -69 26 00.0 & 2008Dec26 & 1.71 & 0.021 \\
..  & .. .. .... & ... .. .... & 2011Dec11 & 1.39 & 0.021 \\
n15 & 05 37 24.2 & -69 25 59.9 & 2008Dec26 & 1.80 & 0.023 \\
..  & .. .. .... & ... .. .... & 2011Dec11 & 1.22 & 0.021 \\
n16 & 05 37 23.7 & -69 32 39.9 & 2008Dec26 & 2.10 & 0.023 \\
..  & .. .. .... & ... .. .... & 2011Dec11 & 1.04 & 0.021 \\
n17 & 05 38 40.2 & -69 32 40.0 & 2011Dec11 & 1.11 & 0.025 \\
n18 & 05 39 56.4 & -69 32 40.1 & 2011Dec11 & 1.07 & 0.025 \\
n19 & 05 39 56.8 & -69 39 20.2 & 2011Dec11 & 1.31 & 0.026 \\
n20 & 05 38 40.1 & -69 39 20.0 & 2011Dec11 & 1.20 & 0.023 \\
n21 & 05 37 23.3 & -69 39 20.0 & 2011Dec11 & 1.26 & 0.022 \\
n22 & 05 37 23.0 & -69 46 00.4 & 2008Dec27 & 1.41 & 0.023 \\
..  & .. .. .... & ... .. .... & 2011Dec11 & 1.26 & 0.024 \\
n23 & 05 38 40.1 & -69 46 00.2 & 2008Dec27 & 1.94 & 0.018 \\
..  & .. .. .... & ... .. .... & 2011Dec11 & 1.15 & 0.026 \\
n24 & 05 39 57.1 & -69 46 00.1 & 2011Dec11 & 1.34 & 0.028 \\
n25 & 05 39 57.5 & -69 52 40.0 & 2011Dec11 & 1.22 & 0.029 \\
n26 & 05 38 40.1 & -69 52 40.3 & 2011Dec10 & 2.18 & 0.024 \\
n27 & 05 37 22.6 & -69 52 40.8 & 2008Dec30 & 1.78 & 0.021 \\
..  & .. .. .... & ... .. .... & 2011Dec10 & 2.05 & 0.025 \\
n28 & 05 36 11.3 & -68 59 19.9 & 2011Dec10 & 1.71 & 0.026 \\
n29 & 05 36 10.7 & -69 06 00.3 & 2011Dec10 & 1.48 & 0.026 \\
n30 & 05 36 09.8 & -69 12 39.9 & 2008Dec25 & 1.74 & 0.027 \\
n31 & 05 36 09.1 & -69 19 20.1 & 2008Dec25 & 1.87 & 0.026 \\
n33 & 05 36 07.4 & -69 32 40.0 & 2008Dec26 & 1.60 & 0.023 \\
n34 & 05 36 06.7 & -69 39 20.1 & 2008Dec27 & 1.63 & 0.022 \\
n35 & 05 36 05.8 & -69 46 00.3 & 2008Dec27 & 1.62 & 0.023 \\
n36 & 05 36 05.5 & -69 52 40.1 & 2011Dec10 & 1.81 & 0.024 \\
n37 & 05 41 08.8 & -68 59 19.9 & 2008Dec30 & 1.64 & 0.024 \\
n38 & 05 41 09.5 & -69 05 59.8 & 2008Dec25 & 1.53 & 0.019 \\
n39 & 05 41 10.4 & -69 12 40.1 & 2008Dec25 & 1.58 & 0.020 \\
n40 & 05 41 11.0 & -69 19 20.1 & 2008Dec25 & 2.17 & 0.019 \\
n44 & 05 41 14.2 & -69 46 00.2 & 2008Dec30 & 1.77 & 0.019 \\
n45 & 05 41 15.0 & -69 52 40.3 & 2008Dec30 & 1.73 & 0.025 \\

\enddata
\tablecomments{Data from 2 to 9 on December 2011 were discarded by instrumental problem during the observation.}
\end{deluxetable}

\clearpage

\begin{landscape}
\begin{deluxetable}{ccccccccccccccccccccccccccccccccc}
\tabletypesize{\tiny}
\setlength{\tabcolsep}{0.02in}
\tablewidth{0pt}
\tablecaption{Photometric and polarimetric catalog of the sources on the northeastern regions in the LMC \tablenotemark{a} \label{tbl-2}}
\tablehead{
\colhead{} & \colhead{} & \multicolumn{2}{c}{Position} &\colhead{}& \multicolumn{6}{c}{Magnitude} &\colhead{}& \multicolumn{12}{c}{Polarization properties} &\colhead{}&
\multicolumn{4}{c}{$SPM$ data} & \colhead{} & \colhead{} & \colhead{} & \colhead{}\\
\cline{3-4} \cline{6-11} \cline{13-24} \cline{26-29}\\
\multicolumn{1}{c}{Field ID} & \multicolumn{1}{c}{Source ID} & \multicolumn{1}{c}{${{\alpha_{\circ\rm J2000.0}}}$} & \multicolumn{1}{c}{${\delta_{\circ\rm J2000.0}}$} & \colhead{}&
\multicolumn{2}{c}{$J$} & \multicolumn{2}{c}{$H$} & \multicolumn{2}{c}{$K_{s}$} & \colhead{} &
\multicolumn{2}{c}{$P_{J}$} & \multicolumn{2}{c}{$P_{H}$} & \multicolumn{2}{c}{$P_{K_{s}}$} &
\multicolumn{2}{c}{$\theta_{J}$} & \multicolumn{2}{c}{$\theta_{H}$} & \multicolumn{2}{c}{$\theta_{K_{s}}$} & \colhead{} &
\multicolumn{2}{c}{$\mu_{\alpha}\cos\delta$} & \multicolumn{2}{c}{$\mu_{\delta}$} &
\colhead{$B$} & \colhead{$V$}\\
\multicolumn{1}{c}{} & \multicolumn{1}{c}{} & \multicolumn{1}{c}{} & \multicolumn{1}{c}{} & \colhead{}&
\multicolumn{2}{c}{(mag)} & \multicolumn{2}{c}{(mag)} & \multicolumn{2}{c}{(mag)} & \colhead{} &
\multicolumn{2}{c}{(\%)} & \multicolumn{2}{c}{(\%)} & \multicolumn{2}{c}{(\%)} &
\multicolumn{2}{c}{($\arcdeg$)} & \multicolumn{2}{c}{($\arcdeg$)} & \multicolumn{2}{c}{($\arcdeg$)} & \colhead{} &
\multicolumn{2}{c}{(mas yr$^{-1}$)} & \multicolumn{2}{c}{(mas yr$^{-1}$)} &
\colhead{(mag)} & \colhead{(mag)}\\
\colhead{(1)}&\colhead{(2)}&\colhead{(3)}&\colhead{(4)}&\colhead{}&\colhead{(5)}&\colhead{(6)}&\colhead{(7)}&\colhead{(8)}&\colhead{(9)}&
\colhead{(10)}&\colhead{}&\colhead{(11)}&\colhead{(12)}&\colhead{(13)}&\colhead{(14)}&\colhead{(15)}&\colhead{(16)}&\colhead{(17)}&\colhead{(18)}&\colhead{(19)}&
\colhead{(20)}&\colhead{(21)}&\colhead{(22)}&\colhead{}&\colhead{(23)}&\colhead{(24)}&\colhead{(25)}&\colhead{(26)}&\colhead{(27)}&\colhead{(28)}}
\startdata
 LMC\_n1\_081230 & 05394434-6902386 & 05 39 44.34 & -69 02 38.53 &&  14.588  &  0.010 &  13.877  &  0.004 &  13.641  &  0.017  &&     $<$  &   2.92  &    $<$  &   1.98  &   5.02  &   1.45  & $\cdots$ & $\cdots$& $\cdots$ &$\cdots$  &   120.01  &     7.93 &&  13.19 & 26.86  &  -8.20 & 25.76 & 18.31 & 18.37 \\
 LMC\_n1\_081230 & 05402245-6855536 & 05 40 22.44 & -68 55 53.55 &&  12.122  &  0.003 &  11.108  &  0.001 &  10.799  &  0.003  &&     $<$  &   0.81  &   0.67  &   0.15  &    $<$  &   0.57  & $\cdots$ & $\cdots$&   134.86 &   6.37 & $\cdots$  & $\cdots$ &&   5.92 & 10.72  &  -0.96 & 10.56 & 18.51 & 16.51 \\
 LMC\_n1\_081230 & 05395766-6855594 & 05 39 57.65 & -68 55 59.29 &&  13.193  &  0.005 &  11.725  &  0.002 &  11.120  &  0.003  &&     $<$  &   1.35  &   0.73  &   0.22  &    $<$  &   0.60  & $\cdots$ & $\cdots$&   144.33 &   8.20 & $\cdots$  & $\cdots$ &&  -9.49 & 26.88  &  -1.86 & 25.79 & 24.42 & 18.37 \\
 LMC\_n1\_081230 & 05402167-6856250 & 05 40 21.66 & -68 56 24.92 &&  14.665  &  0.011 &  13.805  &  0.004 &  13.562  &  0.016  &&     $<$  &   3.03  &   2.08  &   0.63  &    $<$  &   4.03  & $\cdots$ & $\cdots$&   143.81 &   8.28 & $\cdots$  & $\cdots$ && $\cdots$ & $\cdots$ & $\cdots$ & $\cdots$ & $\cdots$ & $\cdots$ \\
 LMC\_n1\_081230 & 05402523-6856327 & 05 40 25.21 & -68 56 32.80 &&  14.384  &  0.009 &  13.413  &  0.003 &  13.133  &  0.012  &&     $<$  &   2.58  &   2.59  &   0.50  &    $<$  &   2.77  & $\cdots$ & $\cdots$&   136.25 &   5.44 & $\cdots$  & $\cdots$ && $\cdots$ & $\cdots$ & $\cdots$ & $\cdots$ & $\cdots$ & $\cdots$ \\
 LMC\_n1\_081230 & 05401186-6856470 & 05 40 11.86 & -68 56 46.89 &&  14.456  &  0.009 &  13.557  &  0.003 &  13.324  &  0.013  &&     $<$  &   2.68  &   2.06  &   0.54  &    $<$  &   3.26  & $\cdots$ & $\cdots$&   122.63 &   7.27 & $\cdots$  & $\cdots$ &&  -2.41 & 18.63  &  34.94 & 17.99 & 18.95 & 17.93 \\
 LMC\_n1\_081230 & 05402353-6856500 & 05 40 23.54 & -68 56 49.85 &&  13.096  &  0.005 &  12.467  &  0.002 &  12.248  &  0.007  &&     $<$  &   1.32  &   1.16  &   0.30  &    $<$  &   1.47  & $\cdots$ & $\cdots$&   135.02 &   7.08 & $\cdots$  & $\cdots$ &&   2.90 &  8.05  &   9.60 &  8.07 & 16.37 & 15.02 \\
 LMC\_n1\_081230 & 05401208-6857259 & 05 40 12.07 & -68 57 25.83 &&  12.555  &  0.004 &  11.610  &  0.001 &  11.348  &  0.004  &&     $<$  &   1.00  &   0.79  &   0.19  &    $<$  &   0.81  & $\cdots$ & $\cdots$&   124.71 &   6.81 & $\cdots$  & $\cdots$ &&   0.52 &  9.38  &   2.91 &  9.33 & 18.08 & 15.64 \\
 LMC\_n1\_081230 & 05392139-6857262 & 05 39 21.40 & -68 57 25.97 &&  14.045  &  0.008 &  13.412  &  0.003 &  13.192  &  0.012  &&     $<$  &   2.15  &    $<$  &   1.49  &   3.39  &   0.99  & $\cdots$ & $\cdots$& $\cdots$ &$\cdots$  &   137.16  &     8.03 &&   2.79 & 11.54  &  19.50 & 11.31 & 17.50 & 16.87 \\
 LMC\_n1\_081230 & 05394537-6857279 & 05 39 45.37 & -68 57 27.95 &&  11.782  &  0.003 &  11.298  &  0.001 &  11.121  &  0.003  &&    0.78  &   0.23  &   0.98  &   0.16  &   1.16  &   0.24  &   148.45 &     8.16&  143.67  &   4.72 &   137.73  &     5.85 &&   2.44 &  5.90  & -11.26 &  5.99 & 13.99 & 13.10 \\
\enddata
\tablecomments{For sources with $P/\sigma_P$ $\le$ 3, the 3$\sigma_P$ upper limits are listed. Information for each column of the table is given below:\\
Column (1) Observation field names that include observation date;\\
Column (2) Source ID; the formats referred to the equatorial coordinates;\\
Column (3)$-$(4) Equatorial coordinates (J2000.0) in decimal degrees;\\
Column (5)$-$(10) $J$, $H$, and $K_{s}$ magnitude and error;\\
Column (11)$-$(16) $J$, $H$, and $K_{s}$ polarization degree and error in units of percentage;\\
Column (17)$-$(22) $J$, $H$, and $K_{s}$ polarization position angle and error in units of degrees;\\
Column (23)$-$(24) Absolute proper motion in right ascension and error in units of mas yr$^{-1}$ from SPM catalog;\\
Column (25)$-$(26) Absolute proper motion in declination and error in units of mas yr$^{-1}$ from SPM catalog;\\
Column (27)$-$(28) Johnson B and V magnitude from SPM catalog.}
\tablenotetext{a}{Only a portion of catalog is listed in Table~\ref{tbl-3}. Complete source catalog is available at  http://irlab.khu.ac.kr/\~{}jaeyeong/Tab3.dat.}
\end{deluxetable}
\end{landscape}

\clearpage

\begin{deluxetable}{ccccccccccccccccc}
\tabletypesize{\scriptsize}
\tablecaption{Statistical Distribution of Polarization Properties at Each Field\label{tbl-3}}
\tablewidth{0pt}
\setlength{\tabcolsep}{0.02in}
\tablehead{\colhead{Field name} & \multicolumn{3}{c}{Number} & \multicolumn{1}{c}{$\langle$$P_{J}\rangle$} & \multicolumn{1}{c}{$\langle\sigma_{J}\rangle$} & \multicolumn{1}{c}{$\langle$$P_{H}\rangle$} & \multicolumn{1}{c}{$\langle\sigma_{H}\rangle$}
& \multicolumn{1}{c}{$\langle$$P_{K_{s}}\rangle$} & \multicolumn{1}{c}{$\langle\sigma_{K_{s}}\rangle$} & \multicolumn{1}{c}{$\langle\theta_{J}\rangle$} & \multicolumn{1}{c}{$\langle\sigma_{J}\rangle$} &
\multicolumn{1}{c}{$\langle\theta_{H}\rangle$} & \multicolumn{1}{c}{$\langle\sigma_{H}\rangle$} & \multicolumn{1}{c}{$\langle\theta_{K_{s}}\rangle$} & \multicolumn{1}{c}{$\langle\sigma_{K_{s}}\rangle$} \\
\cline{2-4}\\
\colhead{} & \colhead{$J$} & \colhead{$H$} & \colhead{$K_{s}$} & \colhead{(\%)} & \colhead{(\%)} & \colhead{(\%)} & \colhead{(\%)} & \colhead{(\%)} & \colhead{(\%)}
& \colhead{($\arcdeg$)} & \colhead{($\arcdeg$)} & \colhead{($\arcdeg$)} & \colhead{($\arcdeg$)} & \colhead{($\arcdeg$)} & \colhead{($\arcdeg$)}\\
\colhead{(1)} & \colhead{(2)} & \colhead{(3)} & \colhead{(4)} & \colhead{(5)} & \colhead{(6)} & \colhead{(7)} & \colhead{(8)} & \colhead{(9)} & \colhead{(10)}
& \colhead{(11)} & \colhead{(12)} & \colhead{(13)} & \colhead{(14)} & \colhead{(15)} & \colhead{(16)}}
\startdata
  n1 &   6 &  15 &   8 &  1.17 &  0.14 &  0.99 &  0.06 &  1.53 &  0.13 &  140.77 &  50.13 &  139.18 &  10.46 &  138.77 &  11.88 \\
  n3 &  31 &  55 &  26 &  1.91 &  0.08 &  1.70 &  0.04 &  1.33 &  0.06 &   57.73 &   7.74 &   63.78 &   8.74 &   62.52 &  11.03 \\
  n4 &  14 &  18 &  10 &  1.47 &  0.11 &  1.30 &  0.08 &  1.86 &  0.18 &   84.18 &  45.32 &   92.96 &  52.09 &   86.87 &  53.53 \\
  n5 &  16 &  18 &  10 &  2.51 &  0.08 &  1.74 &  0.06 &  2.45 &  0.11 &  149.69 &  58.36 &  107.42 &  42.60 &  128.77 &  48.95 \\
  n6 &   4 &   3 &   2 &  1.80 &  0.25 &  1.34 &  0.14 &  1.30 &  0.21 &   78.15 &   7.77 &   75.60 &   6.37 &   83.57 &  18.72 \\
  n7 &  18 &  29 &  15 &  1.58 &  0.09 &  1.24 &  0.05 &  1.25 &  0.08 &   15.65 &  17.26 &    7.09 &  12.66 &    5.79 &  31.04 \\
  n8 &  15 &  22 &  15 &  1.98 &  0.10 &  1.53 &  0.06 &  1.08 &  0.07 &   69.46 &  53.88 &   97.91 &  40.94 &  103.47 &  33.55 \\
  n9 &  30 &  52 &  31 &  2.27 &  0.08 &  1.92 &  0.04 &  2.01 &  0.08 &   91.97 &  18.48 &   93.67 &  16.64 &   95.51 &  19.09 \\
 n10 &  17 &  60 &  29 &  1.86 &  0.09 &  1.67 &  0.04 &  1.87 &  0.09 &   47.35 &   6.67 &   51.58 &   8.09 &   49.39 &  15.60 \\
 n11 &  10 &  15 &   8 &  1.19 &  0.10 &  1.05 &  0.07 &  1.96 &  0.21 &   28.48 &  14.34 &   26.84 &  18.37 &    3.76 &  44.49 \\
 n12 &  28 &  56 &  25 &  1.78 &  0.07 &  1.57 &  0.04 &  1.37 &  0.08 &  134.81 &  13.19 &  130.56 &  22.64 &  116.69 &  36.68 \\
 n13 &  10 &  32 &  17 &  1.98 &  0.12 &  1.20 &  0.05 &  1.54 &  0.10 &  127.88 &  22.76 &  120.76 &  26.63 &  129.26 &  22.04 \\
 n14 &  79 &  99 &  46 &  2.51 &  0.05 &  1.82 &  0.03 &  1.70 &  0.05 &   10.98 &  13.37 &    9.03 &  17.13 &   11.45 &  21.94 \\
 n15 &   5 &   8 &  17 &  0.81 &  0.13 &  1.56 &  0.15 &  1.90 &  0.16 &   10.87 &  29.24 &   29.18 &  21.07 &    3.86 &  41.04 \\
 n16 &   6 &  27 &  13 &  1.05 &  0.11 &  0.89 &  0.04 &  1.15 &  0.11 &   60.02 &  57.63 &   78.98 &  46.74 &   75.73 &  47.31 \\
 n17 &  24 &  26 &  10 &  2.50 &  0.11 &  1.75 &  0.06 &  1.67 &  0.12 &    0.90 &  23.38 &    6.00 &  19.80 &   14.26 &  34.78 \\
 n18 &  13 &  20 &  10 &  1.59 &  0.12 &  1.26 &  0.07 &  1.73 &  0.16 &    1.01 &  32.24 &   15.32 &  33.57 &    6.84 &  19.94 \\
 n19 &  11 &  11 &   7 &  2.17 &  0.13 &  1.85 &  0.08 &  1.34 &  0.12 &   69.68 &  58.42 &   65.08 &  38.02 &   88.81 &  52.04 \\
 n20 &   8 &  15 &   8 &  1.99 &  0.17 &  1.12 &  0.08 &  1.35 &  0.15 &   97.36 &  66.20 &  112.08 &  50.86 &  115.97 &  52.49 \\
 n21 &   3 &   6 &   6 &  3.08 &  0.44 &  1.84 &  0.15 &  3.17 &  0.30 &   59.16 &  53.53 &  112.92 &  64.84 &   78.65 &  47.78 \\
 n22 &  28 &  60 &  36 &  2.00 &  0.09 &  1.34 &  0.04 &  1.43 &  0.06 &   65.76 &  23.22 &   71.22 &  29.12 &   73.70 &  40.13 \\
 n23 &  49 &  80 &  30 &  1.51 &  0.05 &  1.28 &  0.03 &  1.25 &  0.06 &   42.70 &  34.21 &   41.98 &  36.16 &   54.99 &  50.11 \\
 n24 &  14 &  22 &  14 &  3.07 &  0.14 &  1.56 &  0.06 &  1.39 &  0.08 &   80.27 &  43.90 &   64.97 &  40.29 &   58.17 &  36.28 \\
 n25 &  14 &  25 &  14 &  2.10 &  0.12 &  1.11 &  0.05 &  1.75 &  0.06 &  109.20 &  38.52 &   69.88 &  33.28 &   68.03 &  40.19 \\
 n26 &  28 &  53 &  35 &  2.10 &  0.09 &  1.59 &  0.04 &  1.45 &  0.05 &   65.35 &  11.71 &   71.07 &   8.46 &   69.85 &  16.67 \\
 n27 &  21 &  49 &  27 &  1.79 &  0.09 &  1.24 &  0.04 &  1.05 &  0.06 &   65.37 &  21.36 &   74.80 &  24.57 &   72.86 &  40.19 \\
 n28 &  28 &  57 &  25 &  2.17 &  0.10 &  1.87 &  0.04 &  1.67 &  0.07 &  115.04 &  17.10 &  104.52 &  13.30 &  114.09 &  15.46 \\
 n29 &   8 &  22 &  10 &  1.49 &  0.14 &  1.15 &  0.06 &  1.03 &  0.10 &   82.35 &  17.27 &   82.22 &  25.71 &   83.08 &  40.00 \\
 n30 &  23 &  33 &  18 &  2.24 &  0.09 &  2.13 &  0.05 &  1.87 &  0.10 &   76.58 &  19.05 &   82.81 &  13.06 &   90.63 &  30.38 \\
 n31 &   7 &  25 &   2 &  1.21 &  0.15 &  1.27 &  0.06 &  3.23 &  0.71 &   64.53 &  18.35 &   73.36 &  13.93 &   65.86 &   4.60 \\
 n33 &   2 &  12 &   8 &  1.10 &  0.28 &  0.98 &  0.07 &  2.25 &  0.20 &  155.01 &   9.73 &  107.10 &  32.60 &   95.92 &  35.90 \\
 n34 &  22 &  29 &  15 &  1.42 &  0.08 &  0.96 &  0.04 &  1.19 &  0.09 &   91.48 &  35.04 &   97.69 &  24.22 &   93.37 &  37.09 \\
 n35 &  16 &  36 &  10 &  1.34 &  0.09 &  1.06 &  0.04 &  0.92 &  0.08 &   69.54 &  16.82 &   79.05 &  13.93 &   75.85 &  14.42 \\
 n36 &  52 &  69 &  29 &  2.10 &  0.07 &  1.58 &  0.04 &  1.36 &  0.06 &   34.89 &  14.35 &   33.31 &  11.71 &   43.49 &  45.81 \\
 n37 &   6 &  19 &   5 &  2.73 &  0.26 &  1.77 &  0.09 &  2.61 &  0.34 &   69.02 &   9.05 &   77.33 &  12.33 &   83.17 &  16.88 \\
 n38 &  47 &  67 &  34 &  2.47 &  0.06 &  2.22 &  0.03 &  1.88 &  0.06 &   77.69 &   5.84 &   77.79 &   6.04 &   75.34 &  13.16 \\
 n39 &  17 &  31 &  12 &  1.32 &  0.08 &  1.11 &  0.04 &  1.01 &  0.08 &    0.59 &  38.90 &    0.71 &  26.83 &    6.30 &  24.68 \\
 n40 &  59 &  68 &  34 &  3.21 &  0.08 &  2.81 &  0.05 &  2.88 &  0.09 &   71.11 &   9.61 &   75.28 &   7.41 &   73.34 &  12.89 \\
 n44 &  31 &  43 &  22 &  1.64 &  0.08 &  1.52 &  0.04 &  1.40 &  0.07 &  159.21 &  29.87 &  155.54 &  41.15 &  155.62 &  25.22 \\
 n45 &  55 &  81 &  39 &  2.13 &  0.06 &  1.74 &  0.03 &  1.41 &  0.06 &    1.98 &   9.26 &    1.20 &  10.88 &    1.40 &  20.00 \\
\enddata
\tablecomments{Information for each column of the table is given below:\\
Column (1) Observation field names;\\
Column (2)$-$(4) the number of stars with $P/\sigma_P$ $>$ 3 for each band;\\
Column (5), (7), (9) The average polarization degrees for each band;\\
Column (6), (8), (10) The mean uncertainties of polarization degrees for each band;\\
Column (11), (13), (15) The average polarization position angles for each band;\\
Column (12), (14), (16) The standard deviation of polarization position angles for each band.}
\end{deluxetable}

\clearpage

\begin{deluxetable}{ccccccc}
\tabletypesize{\scriptsize}
\tablecaption{Magnetic Field Properties at Sample Fields\label{tbl-4}}
\tablewidth{0pt}
\setlength{\tabcolsep}{0.02in}
\tablehead{\colhead{} & \colhead{Number of } & \colhead{$\delta\textit{v}_{los}$} & \colhead{$\delta\theta$} & \colhead{Normality } & \colhead{$B$} & \colhead{$B_{W07}$\tablenotemark{a} } \\
\colhead{Field name} & \colhead{samples} & \colhead{(km s$^{-1}$)} & \colhead{($\theta$)} &\colhead{test value} & \colhead{($\mu$G)} & \colhead{($\mu$G)}}
\startdata
n8 & 22 & 1.4 & 15 & 0.92 & 3 & \\
n26 & 54 & 8.0 & 9 & 0.95 & 25 & \\
n29 & 22 & 1.5 & 16 & 1.42 & 3 & \\
n30 & 33 & 3.3 & 13 & 1.02 & 7 & \\
n40 & 69 & 4.8 & 8 & 1.01 & 19 & 30\tablenotemark{b} \\
\enddata
\tablenotetext{a}{Magnetic field strength from W07}
\tablenotetext{b}{Region D}
\end{deluxetable}

\clearpage

\clearpage

\clearpage

\clearpage

\clearpage

\clearpage

\clearpage

\clearpage

\clearpage

\clearpage

\end{document}